\documentclass[9pt,twocolumn]{article}

\usepackage{authblk}
\usepackage{graphicx}
\usepackage{float}
\usepackage{caption}
\usepackage{siunitx}
\usepackage{textcomp}
\usepackage{balance}
\usepackage[runin]{abstract}

\usepackage{amsmath,amssymb,latexsym}
\usepackage{subcaption}
\begin{document}
\title{  Frequency comb based four-wave-mixing  spectroscopy}

\author[1,2]{Bachana Lomsadze}
\author[1,2]{Steven T. Cundiff}

\affil[1]{JILA, University of Colorado \& National Institute of Standards and Technology, Boulder, Colorado 80309-0440, USA}

\affil[2]{Department of Physics, University of Michigan, Ann Arbor, Michigan 48109, USA}

\affil[*]{Corresponding author: cundiff@umich.edu} 

\setlength{\columnsep}{1.2cm}
\twocolumn[
 \date{}
\maketitle




We experimentally demonstrate  four-wave-mixing spectroscopy using frequency combs. The experiment uses a geometry where excitation pulses and  four-wave-mixing signals generated  by  a sample  co-propagate. We separate them in the radio frequency domain by heterodyne detection with a local oscillator comb that has a different repetition frequency.
\vspace*{2 cm}

]



The precision and sensitivity of optical spectroscopy makes it a powerful tool for both fundamental science, such as the historic observations of discrete spectral lines that provided evidence for the quantization of atomic energy levels, and for applications such as chemical sensing and materials characterization. Nevertheless, the standard methods, which work in the linear regime, have significant limitations when faced with inhomogeneously broadened transitions or heterogeneous samples. Furthermore, increasing the spectral resolution requires a commensurate increase in the physical size of the spectrometer, and/or the acquisition time, which is incompatible with many potential applications.

The development of dual comb spectroscopy (DCS) ~\cite{Keilmann:04, newburyPRL} has provided a route to simultaneously achieving high resolution and short acquisition times using a compact apparatus. These attributes have led to DCS being demonstrated for many applications ~\cite{Coddington:16, rieker2014frequency, Godbout:10}. However DCS is a linear method and thus cannot address the issues of inhomogeneously broadened transitions or sample heterogeneity.

 \begin{figure*}[!]
\centering
\includegraphics[width=0.8\linewidth]{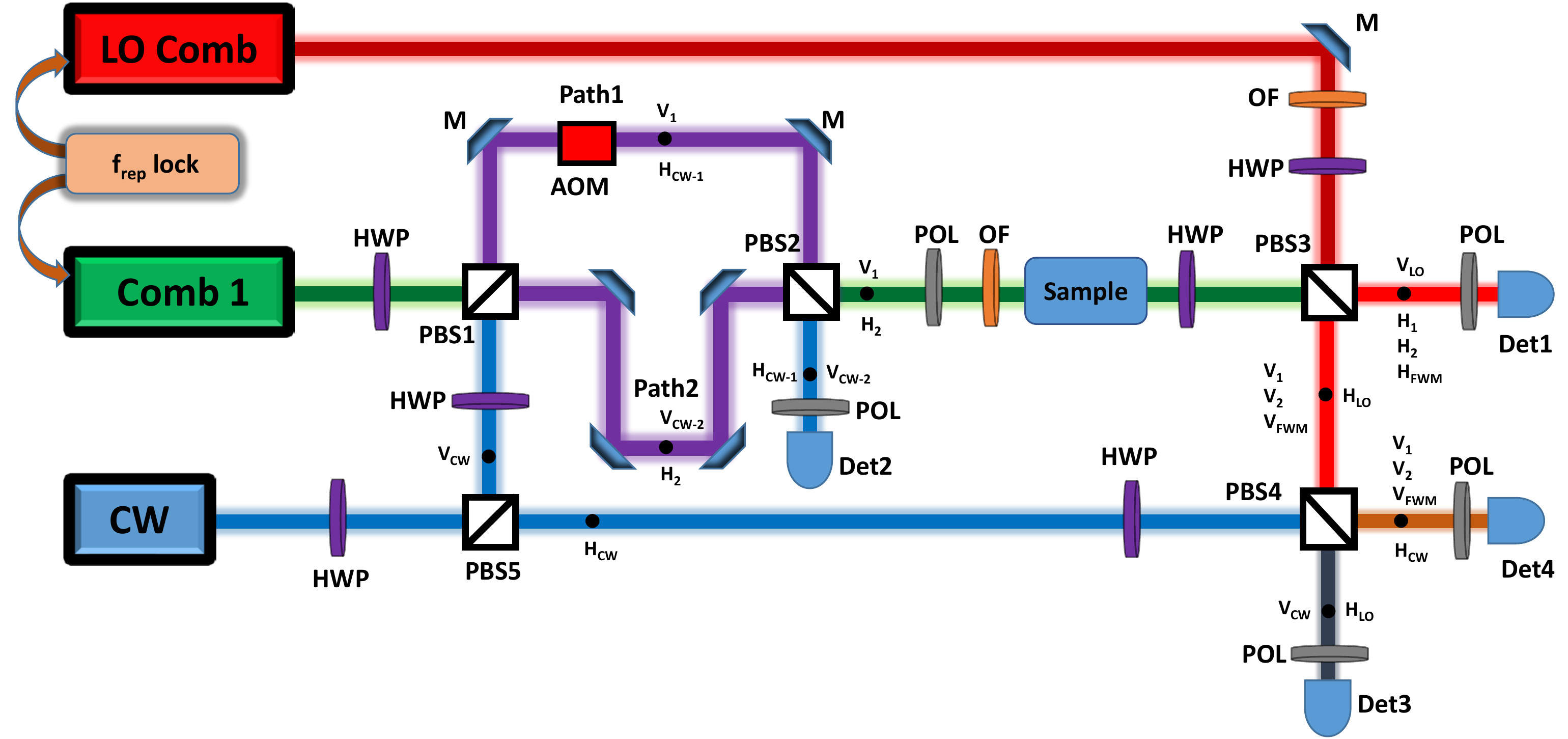}
\caption{  Experimental Setup: CW-Continuous Wave laser, LO-Local oscillator, FWM-Four-Wave-Mixing,  HWP-Half Wave Plate, PBS-Polarizing Beam Splitter, AOM-Acousto Optical Modulator, M-Mirror, POL-Polarizer, OF-Optical Filter, BPF-RF Band Pass Filter, Det-detector.
H -Horizontal and V-Vertical indicate the linear polarization states of the beams. Subscripts 1 and 2 indicate the path of the beams. }
\label{fig:experiment}
\end{figure*}

To overcome the limitations of linear spectroscopy, nonlinear methods must be used. Many variants of nonlinear spectroscopy have been developed, each with its own advantages and area of utility. Many of them rely on some form of four-wave-mixing, a fundamental nonlinear optical process.

Four-wave-mixing spectroscopy encompasses many different methods including Raman Induced Kerr Effect Spectroscopy (RIKES), Coherent Anti-stokes Raman Spectroscopy (CARS), Stimulated Raman gain spectroscopy ~\cite{boyd,levenson2012introduction, Agarwal} and photon echo spectroscopy ~\cite{photonecho}. These methods have proven to be very powerful for chemical sensing, optical imaging of biological tissues and studying chemical reactions \cite{SunneyXie}. FWM underpins optical multidimensional coherent spectroscopy (MDCS), which is frequently used to study complex  properties and ultrafast dynamics of a wide range of materials ~\cite{tokmakoff, hamm2011concepts, gael2d, Cundiff:12, schlau2012ultrafast}. These methods however have limited spectral resolution. In addition, most of these techniques have long acquisition times and utilize complex geometries and/or phase cycling schemes to suppress the background signals.

To leverage the advantages offered by frequency comb technology, we demonstrate four-wave-mixing (FWM) spectroscopy using an approach that is inspired by DCS. Our approach can simultaneously provide high resolution and short acquisition times, can be implemented in a compact apparatus, is not limited by inhomogeneous broadening, and can determine if resonances are from the same species or different species in a heterogeneous sample.

~ To realize a general approach to frequency comb based four-wave mixing spectroscopy, we use a geometry in which the excitation pulses can be co-linear and have the same linear polarization state. In this geometry instead of suppressing the linear signals we simultaneously acquire both linear and FWM signals generated by the sample. Even when the optical spectra of these signals overlap (as is the case for this experiment) we are able to separate these contributions.  The separation is performed in the RF domain after heterodyne detection with a local oscillator comb that has a slightly different repetition frequency. By contrast, implementations of CARS and RIKES based on dual-comb spectroscopy  \cite{Ideguchi:12, ideguchi2013coherent}  separate the generated FWM and linear background signals in the optical domain by relying on the magnitude of Raman shifts or choosing the polarization of the incident beams.

~We also present a scheme to subtract the phase noise due to frequency and the optical path fluctuations from the FWM signal in real time. 

~A schematic diagram of the experimental setup is shown in Fig.~\ref{fig:experiment}.
We used two home-built Kerr-lens mode-locked Ti:Sapphire lasers centered at 800 nm. The repetition frequencies of the combs ($f_{rep_{1}} $=93.543954 MHz for comb 1 and $f_{rep_{lo}} $=93.543954 MHz+200.2 Hz for the LO comb), were phase locked to a direct digital synthesizer using a feedback loop. The comb offset frequencies were not stabilized. The output of comb 1 was split into two parts using a half wave plate (HWP) and a polarizing beam splitter (PBS 1). The offset frequency of the first part was shifted by an Acousto-Optical Modulator (AOM) and recombined with the second part on PBS 2. Optical paths for the two arms were adjusted to overlap the two pulse trains in time. Before interacting with the sample, the beams were projected on the same linear polarization state using a polarizer. The sample used for this proof-of-concept experiment was 10 layers of 10 nm GaAs quantum wells (QW) separated by 10 nm thick Al$_{0.3}$Ga$_{0.7}$As barriers. The sample was cooled down to 7 K. The laser beams were optically filtered to excite only the Heavy Hole (HH) excitonic resonance, attenuated and focused to 30 \si{\micro\m} spot on the sample. Average powers for beams traveling through path 1 and 2  were 300 and 600 \si{\micro\W} respectively. The FWM signals emitted from the sample, along with the incident beams, were combined with the LO comb (optically filtered to match filtered comb 1) on PBS 3. Half wave plates were adjusted such that the most of the light from each beam was sent to a photodetector (Det 1) to obtain a RF comb spectrum whereas only small fractions were sent to PBS 4 for phase cancellation (described below).

~The output of Det 1 contains both linear and FWM RF signals, which are spectrally separated in the RF domain. 
\begin{figure}[h!]
\includegraphics[width=\linewidth]{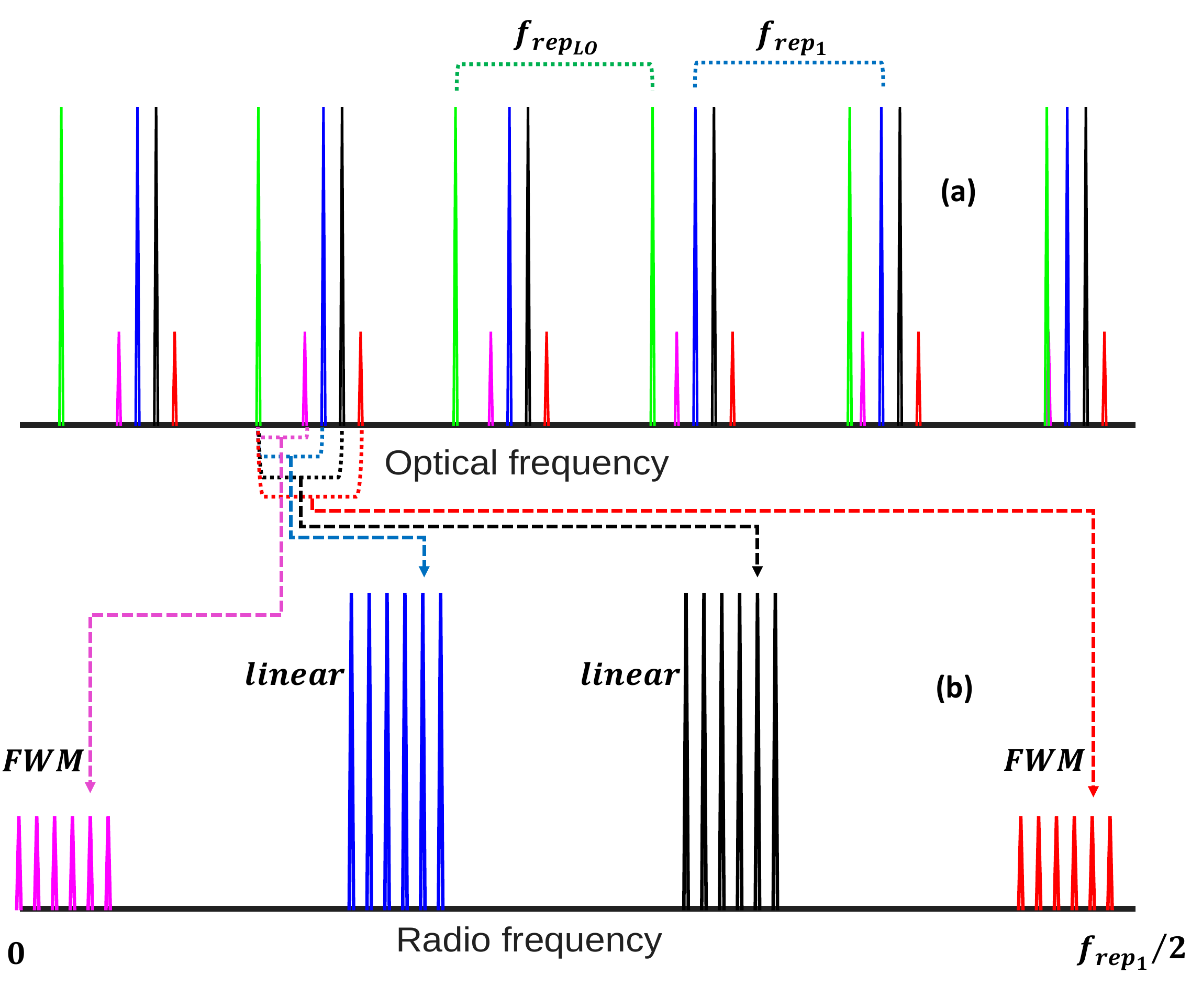}

\centering
\caption{ (a) - Schematic diagram showing separation of  linear and FWM dual comb signals: Black lines-comb1, blue lines- AOM shifted comb 1, red and magenta lines-FWM combs, green lines-LO comb. $f_{rep_{LO}} $ spacing between LO comb teeth, $f_{rep_{1}} $ spacing between comb1 teeth.  (b) - down converted RF frequencies in 0 to half repetition frequency range.}
\label{fig:comb_teeth}
\end{figure}

The separation of linear and FWM RF signals is depicted in Fig.~\ref{fig:comb_teeth}. In Fig.~\ref{fig:comb_teeth} (a) the black and blue lines correspond to the original and AOM shifted comb 1 teeth with $f^{n}_{2}$=$n$$f_{rep_{1}}$+$f_{0_{1}}$ and $f^{n}_{1}$=$n$$f_{rep_{1}}$+$f_{0_{1}}$+$f_{AOM}$ frequencies respectively, where $f_{rep_{1}}$ and $f_{0_{1}}$ are the repetition and offset frequencies of comb 1, $f_{AOM}$ is the AOM drive frequency, and $n$ is an integer. The red and magenta lines correspond to the teeth of the FWM combs with frequencies (-$f^{n}_{1}$ + $f^{m}_{2}$+$f^{k}_{2}$ ) and (-$f^{n}_{2}$ + $f^{m}_{1}$+$f^{k}_{1}$ ) generated by FWM in the sample. Both FWM combs are present only at zero delay but only one is generated for finite delay due to causality \cite{causality}. The green lines with frequencies $f^{n}_{LO}$=$n$$f_{rep_{LO}}$+$f_{0_{LO}}$ represent the teeth of the LO comb, where $f_{rep_{LO}}$ and $f_{0_{LO}}$ are the repetition and offset frequencies of the LO comb. Figure~\ref{fig:comb_teeth} (b) shows the corresponding RF spectrum that shows the clear separation of linear and FWM dual comb signals.

 In the experiment, we isolate one of the FWM signals using a RF bandpass filter before digitizing.

~To obtain a comb structure, the phase fluctuations (described below) of the FWM RF comb teeth  have to be monitored and corrected. The phase of a FWM RF comb tooth (e.g. the red comb on Fig.~ \ref{fig:comb_teeth}) is

\begin{equation} 
\phi^{RF}_{FWM} = (\phi_{1}-\phi_{2})+(\phi_{LO}-\phi_{2})
\label{eq:phase}
\end{equation}
where $\phi_{2}$,  $\phi_{1}$ and $\phi_{LO}$ are the phases of the original, AOM frequency shifted comb 1 and  LO comb teeth respectively. Any uncorrelated fluctuations of these phases will broaden the FWM comb teeth in the RF spectrum. In our co-linear experimental arrangement, the broadening could be due to the fluctuating offset frequencies, residual relative repetition frequency fluctuations and optical path fluctuations between path 1 and path 2. These fluctuations will affect the ($\phi_{LO}$-$\phi_{2}$) and ($\phi_{1}$-$\phi_{2}$) terms in equation~\ref{eq:phase} respectively.

~In the experiment, these fluctuations were measured using an external cavity diode continuous wave (CW) laser (shown in Fig.~\ref{fig:experiment}) that had a wavelength tuned near the HH resonance. The CW laser beam was split into two parts using a HWP and PBS 5. One part of the beam was used to monitor optical path fluctuations between path 1 and path 2. It was split again using a HWP and PBS 1. These beams  propagated through path 1 and 2,  combined on PBS 2 and a beat signal was measured  on a  photodetector (Det 2). The second part of the CW laser beam was used to determine the relative offset frequencies, including fluctuations, of comb 1 and LO comb. It was split on PBS 4, and  interfered with the LO comb and comb 1 on Det 3 and Det 4 photodetectors, respectively. The beat signal on Det 2  that measures the relative path fluctuations between path 1 and path 2 and is also equivalent to measuring the relative phase ($\phi_{2}^{'}-\phi_{1}^{'})$, where $ \phi_{2}^{'}$ and $\phi_{1}^{'}$ correspond to the phases of  the original and the AOM shifted comb 1 teeth nearest to the CW laser frequency. The beat signals measured by Det 3 and Det 4  reflect the optical frequency fluctuations (due to  offset and residual repetition frequency fluctuations) of comb 1 and LO comb teeth nearest to the CW laser frequency: ($\phi_{2}^{'}-\phi_{CW}^{'})$,  ($\phi_{1}^{'}-\phi_{CW}^{'})$ and ($\phi_{LO}^{'}-\phi_{CW}^{'})$, where ($\phi_{CW}^{'})$ is the phase of the CW laser and ($\phi_{LO}^{'})$ is the phase of the LO comb tooth nearest to the CW laser frequency.

\begin{figure}[t!]
\centering
\includegraphics[width= 1.0\linewidth]{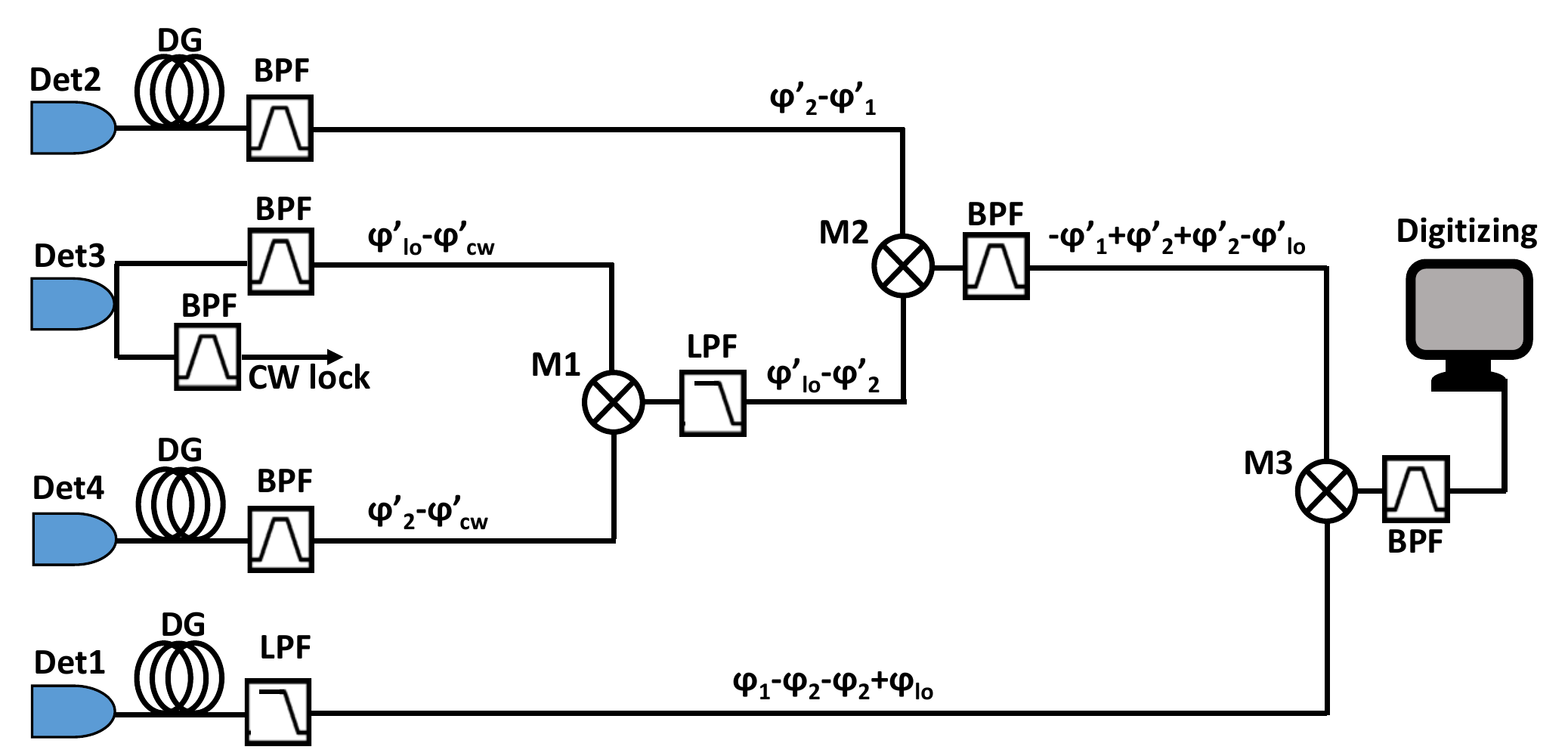}
\caption{Schematic diagram of the phase cancellation scheme: $\phi_{2}^{'}$, $\phi_{1}^{'}$ and $\phi_{LO}^{'}$  correspond to phases of comb 1, AOM shifted comb 1 and LO comb teeth nearest to CW laser frequency. $\phi_{CW}^{'}$ is the phase of CW laser. Det-Detector,  LPF-Low Pass Filter, BPF-Band Pass Filter, DG-delay generator, M-Mixer.  }
\label{fig:phase_cancel}
\end{figure}

~We used these beat  signals to generate a correction signal in real time. Fig.~\ref{fig:phase_cancel} shows a diagram of the phase cancellation scheme.
The beat signal from Det 3 was split into two parts. One part of the signal was bandpass filtered and used for stabilizing the CW laser frequency using a slow loop filter \cite{CWlocking}. 
The second part of Det 3 signal was mixed with the beat signal between the CW laser and a comb 1 tooth (travelling through path 2) from Det 4 that was isolated using a RF band pass filter.
The output of the mixer 1 was filtered in the frequency domain to isolate the signal that had the common CW laser noise canceled and mixed with the signal from Det~2. The output of mixer 2 was band pass filtered again to isolate the signal corresponding to a tooth of the FWM RF comb. Finally, this reference was mixed with the FWM RF comb, filtered and the phase noise free signal was digitized.

 All optical delays as well as the delays introduced by RF cables and filters were measured separately and compensated using the delay generators.

~The phase cancelation scheme treats  $f_{rep}$ fluctuations as  $f_{0}$ fluctuations, which causes the cancelation to degrade for teeth that are far away from the tooth measured by the CW laser. Thus the bandwidth for acceptable phase cancellation  strongly depends on how tightly the repetition frequencies are locked.

\begin{figure}[t!]
\centering
\includegraphics[width=1\linewidth]{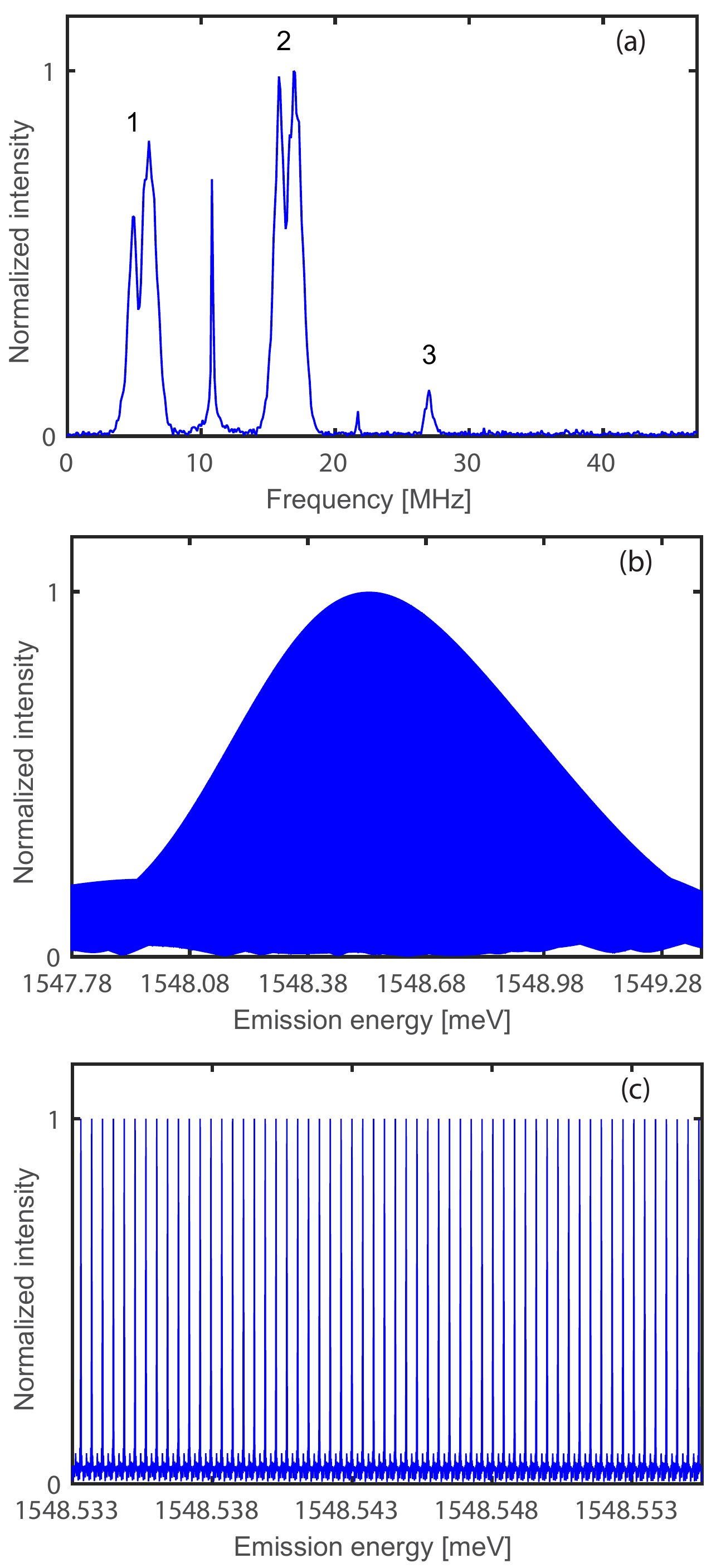}
\caption{(a) Fourier transform of one $^{"}$burst$^{"}$. Peak 1 and 2 correspond to linear signals, 3 corresponds to  FWM signal. (b) Fourier transform of filtered FWM time domain signal (25 $^{"}$bursts$^{"}$) . (c) portion of figure (b) on an expanded scale.}
\label{fig:2}
\end{figure}

 Figure~\ref{fig:2} presents the experimental results. The Fourier transform of one “burst” (corresponding to the temporal overlap of the linear and FWM signals with LO pulses) is shown in Fig.~\ref{fig:2} (a). Peaks 1 and 2 correspond to the linear dual comb signals, which are the result of the heterodyne detection of the AOM shifted comb 1 and the original comb 1 with the LO comb, respectively. The dips are due to the linear absorption of the GaAs quantum well. Peak 3 corresponds to the heterodyne beat between the FWM signal generated by the sample and the LO comb. The figure clearly shows the separation of the linear and FWM contributions. In this figure, only one FWM signal is shown. This is because the pulses traveling through path 2 were slightly delayed in time with respect to the pulses traveling through path 1 ( Fig.~\ref{fig:experiment}). In this case only one FWM signal is generated due to causality \cite{causality}. The spike around 11 MHz corresponds to a beat signal between the original and the AOM frequency shifted combs.

After filtering the FWM RF signal 
the Fourier transform of 25 bursts is shown in  Fig.~\ref{fig:2} (b) (frequency remapped to the optical domain \cite{remap}) and a portion is shown in Fig.~\ref{fig:2} (c) on an expanded scale. The measured spectrum clearly shows the comb structure. The measured linewidth of the FWM signal, corresponds to inhomogeneous broadening of QW, is in good agreement with the literature value \cite{HHlinewidth}.

In conclusion, we have proposed and experimentally demonstrated   four-wave-mixing spectroscopy based on frequency combs. We showed how linear and FWM signal contributions can be separated even when their optical spectra are overlapped and  the excitation pulses have co-linear geometry and the same polarization state. We also demonstrated  a phase cancellation scheme that allowed us to obtain a FWM signal with clear comb structure in the frequency domain even when the combs were not tightly phase locked  and propagated through different optics.

This method  can potentially be used for high resolution, background free spectra- imaging  as well as can be applied to multi-dimensional spectroscopy \cite{stevemukamel} to improve resolution and acquisition speed.
\balance

\section*{Funding Information}
Intelligence Advanced Research Projects Activity (IARPA). Contract  2016-16041300005.


\begin{thebibliography}{10}
\bibitem{CWlocking}
In our scheme the fluctuations of the CW laser were passively canceled (shown
  in Fig.~\ref{fig:phase_cancel}), and the only reason we locked the frequency
  was to avoid slow drifts of the beat signals from active regions of the RF
  band pass filter. Beat signal drifts due to offset frequency drifts were very
  small (less than 10 MHz in 1 hour) and they were measured separately using
  another CW laser that was locked to a Rb atomic hyperfine line.
\bibitem{remap}
For this proof of concept experiment the goal was not to determine the absolute
  optical frequency of the FWM signal. The absolute frequency of the CW laser
  that served as a marker for frequency remapping from the RF to optical domain
  was measured with a low resolution spectrometer.
\end{thebibliography}
\end{document}